# Ferroelectric domain triggers the charge modulation in semiconductors

Anna N. Morozovska,[1,2] Eugene A. Eliseev,[3] Anton V. Ievlev,[4,5] Olexander V. Varenyk[6a], Anastasiia S. Pusenkova[6b], Ying-Hao Chu[7,8], Vladimir Ya. Shur[5*], Maksym V. Strikha [2†], and Sergei V. Kalinin[4‡,]

[1] Institute of Physics, National Academy of Sciences of Ukraine,
46, pr. Nauky, 03028 Kyiv, Ukraine

[2] Institute of Semiconductor Physics, National Academy of Sciences of Ukraine,
41, pr. Nauky, 03028 Kyiv, Ukraine

[3] Institute for Problems of Materials Science, National Academy of Sciences of Ukraine,
3, Krjijanovskogo, 03142 Kyiv, Ukraine

[4] The Center for Nanophase Materials Sciences, Oak Ridge National Laboratory,
Oak Ridge, TN 37831

[5] Ural Federal University, 51, Lenin Ave, 620000 Ekaterinburg, Russia

[6] Taras Shevchenko Kyiv National University, Radiophysical Faculty[a], Physics Faculty[b],
4, pr. Akademika Hlushkova, 03022 Kyiv, Ukraine

[7] Department of Material Science and Engineering, National Chiao-Tung University,
Hsinchu, Taiwan

[8] Institute of Physics, Academia Sinica, Taipei 105, Taiwan

We consider a typical heterostructure "domain patterned ferroelectric film − ultra-thin dielectric layer − semiconductor", where the semiconductor can be an electrolyte, paraelectric or multi-layered graphene. Unexpectedly we have found that the space charge modulation profile and amplitude in the semiconductor, that screens the spontaneous polarization of a 180-degree domain structure of ferroelectric, depends on the domain structure period, dielectric layer thickness and semiconductor screening radius in a rather non-trivial nonlinear way. Multiple size effects appearance and manifestation are defined by the relationship between these three parameters. In addition, we show that the concept of effective gap can be introduced in a simple way only for a single-domain limit. Obtained analytical results open the way for understanding of current-AFM maps of contaminated ferroelectric surfaces in ambient atmosphere as well as explore the possibilities of conductivity control in ultra-thin semiconductor layers.

[*] vladimir.shur@urfu.ru
[†] maksym_strikha@hotmail.com
[‡] sergei2@ornl.gov

# 1. Introduction

Ferroelectric single-crystals, thin films and ferroelectric-based nanosized heterostructures attract permanent interest due to their unique electrophysical and electromechanical properties and promising possibilities for next-generation of memory [1, 2, 3, 4, 5, 6] and nonlinear optical devices [7, 8]. Study of the polarization switching keeps central place in the investigations of ferroelectric materials.

In the general case switching process is a completion between external electric field, depolarization field produced by the boundary charges on the polar surfaces and charged domain walls and screening fields produced by the redistributed charges in the external electrodes (*external screening*) and in the crystal bulk (*bulk screening*) [9, 10, 11]. In that way screening plays important role in the switching process.

Relaxation time of the external screening is defined by characteristics of the external circuit and usually is about microseconds while typical time of the bulk screening can exceed hundreds of milliseconds. Existing of the effective dielectric layer on the sample surfaces [9, 10] leads to spatial separation of the boundary and screening charges and thus essentially inhomogeneous distribution of the electric field in vicinity of domain wall. Recent experimental investigations of the polarization switching in homogeneous electric field in lithium niobate and lithium tantalite single-crystals demonstrated that this phenomenon could lead to unexpected domains behavior like discreet switching [10]. Experiments with artificially thick dielectric gap produced by Ar ions implantation confirmed influence of the dielectric layer thickness on polarization reversal kinetics [12].

Hence careful theoretical calculations of the spatial distribution of electric fields in vicinity of fresh (not completely screened by bulk processes) domain wall is required to describe polarization process in details. In this way multilayer ferroelectric heterostructures play the role of suitable model object.

Recently Tra-Vu Than et al. [13] demonstrated a generic approach to use a functional layer, ferroelectric thin film $Pb(Zr_{0.2}Ti_{0.8})O_3$ deposited on a protective dielectric layer $LaAlO_3$ followed by a $SrTiO_3$, as a nonvolatile modulation of the electric transport and reported about the colossal change of conduction and metal-insulator transition. At that the spontaneous polarization direction in ferroelectric determines the free carrier type in a semiconductor. Motivated by the finding we decided to study theoretically how the ferroelectric domain structure triggers the space-charge modulation in semiconductors layers of different nature. As typical example we chose the hetero-structure "semiconductor/ultra-thin dielectric layer/ferroelectric film with domain structure". As semiconductors we consider an *electrolyte*, *paraelectric Sr(Ru,Ti)O$_3$*, or *multi-layer graphene*, dielectric passive layers (spontaneous and artificial), as ferroelectrics – *LiNbO$_3$* with relatively high spontaneous polarization and very high coercive field and *NaKC$_4$H$_4$O$_6$·4H$_2$O* (*Rochelle salt*) with relatively low polarization and field, because the structures of such type present a significant interest for different applications as argued below.

Specially deposited dielectric layers (typically polymers, high-k dielectrics or paraelectrics) serve as protective buffer layer between the ferroelectric interface and semiconductor/graphene [13]. However when not created specially, dielectric ("dead" or "passive") layers often appear spontaneously on a ferroelectric surface and substantially influence on the spontaneous polarization screening by external screening charges and mobile adsorbates [10, 14, 15, 16, 17, 18]. In particular a water molecules condensate on a hydrophilic surfaces of ferroelectrics LiNbO$_3$, (Ba,Sr)TiO$_3$, (Pb,Zr)TiO$_3$ and NaKC$_4$H$_4$O$_6$·4H$_2$O placed in ambient conditions, and becomes electrolyte after dissociating on H$^+$ and OH$^-$. For the case ferroelectric polarization and the screening (ambient) charges are separated by the dead layer; the charges dynamics sometimes leads to changing of the switching kinetics [19] and fascinating phenomena such as intermittency, quasiperiodicity and chaos in probe-induced ferroelectric domain switching in the system humid air/dead layer/LiNbO$_3$ [20]. Note, that the lack of theory precludes understanding of complex current-AFM measurements results obtained for contaminated ferroelectric surfaces in ambient atmosphere with moderate humidity.

The theory of the dead layer impact on the ferroelectric properties was firstly proposed by Tagantsev et al (see Ref.[21, 22] and refs. therein). Assuming the existence of a thin layer of nonswitchable dielectric material (whose dielectric constant is much less than that of the ferroelectric) connected in series with the ferroelectric Tagantsev et al. [21] modeled polarization hysteresis loops corresponding to a heterostructure electrode/passive dead layer/ferroelectric/electrode. They predicted that the increase of the passive layer thickness results in an essential tilt of the loops, an essential reduction of the remanent polarization and to a certain reduction of the maximal polarization on the loop and the coercive field. Then Abe et al.[23] used non-switchable layer model to explain the voltage offset of ferroelectric loops.

Since a single layer graphene was fabricated in 2004 [24], numerous theoretical and experimental works dealt with it's unique properties (see e.g. Rev. [25] and refs therein). However, in fact multi-layer graphene (or ultra-thin graphite-like) films are observed in real situations rather often. Therefore electric transport [26, 27], magnetic properties and magneto-optical conductivity [26, 28], thermal conductivity [29] and other characteristics of multi-layer graphene films were studied rather intensively in last years. Note, that, despite single-layer graphene is a semi-metal with zero gap between conduction and valence bands in Dirac point (see e.g. [30], a potential difference between graphene layers opens a gap between conduction and valence bands; in bilayer graphene such a gap can be caused by electric field of gate doping [31], by tensional strain [32, 33], by a twist between two layers in Bernal stacking [34, 35]. Therefore generally a distorted multi-layer graphene can be treated as a semiconductor with controlled gap [36], and it can be modeled very naturally in the problem below along with such a typical semiconductor as silicon.

Several experimental and theoretical studies consider intriguing physical properties of the heterostructure graphene/physical gap or dielectric layer/ferroelectric substrate [37, 38, 39, 40, 41, 42, 43]. Interest to the heterostructures is primary related with tempting possibility to add next level of functionality by electric field and temperature control over the spontaneous polarization direction, value and domain structure properties in the vicinity of surface [43]. A strong depolarization electric field, caused by the abrupt of spontaneous polarization at the ferroelectric surface, partially drops in the dielectric gap and then is screened by the carriers localized in graphene. The charge of the screening carriers is defined by the spontaneous polarization direction [43].

This work presents a comprehensive continuum media theory of the ferroelectric domain structure influence on the space-charge modulation in different semiconductor surface layers allowing for the presence of dielectric layers on the ferroelectric surface. The paper is organized as following. The problem statement, electrostatic equations for the hetero-structure semiconductor/ultra-thin dielectric layer/ferroelectric film with domain structure and their analytical solution are listed in the Section 2 and Appendixes. The electric field and potential redistribution and space charge modulation caused by ferroelectric domain structure are discussed in the Section 3. The finite size effects of the space charge accumulation in semiconductor caused by a ferroelectric domain structure are analyzed in the Section 4. Section 5 is concluding remarks.

## 2. Problem statement

The screening of electric-field in semiconductor/multi-layer graphene can be treated either within Debye-Hückel approximation (for non-degenerated carriers, that obey Maxwell-Boltzmann statistics), or within Fermi-Thomas approximation (for degenerated carriers, that obey Fermi-Dirac statistics, see e.g. [44]). The realistic values for the screening length are in the range between tens nm (non-degenerated silicon) and 1-3 nm (strongly gated multi-layer graphene). In this work, however, we'll study the Debye-Hückel approximation case only, leaving the Fermi-Thomas limit for our further examination. Therefore we restrict ourselves to the case of multi-layer graphene with low gate/ferroelectric doping, where intrinsic carriers 2D concentration exceed one, caused by external field doping. This means we need ferroelectric with comparatively low spontaneous polarization for validity of Debye-Hückel approximation for multi-layer graphene. We also treat multi-layer graphene as a 3D system, characterized by graphite parameters (permittivity $\varepsilon_S$ =15). This imposes limitations on the number of graphene layers N > 10. Debye-Hückel approximation is quite common for typical semiconductor-paraelectrics, solid or liquid electrolytes.

Geometry of the considered problem is shown in the **Figure 1.** Ultra-thin dielectric layer (e.g. physical gap or contaminated layer) is modeled by a dielectric layer of thickness $h$; $l = L - h$ is the thickness of ferroelectric film with 180-degree domain structure. The period of 180-degree domain structure is $a$. $\mathbf{P}_S = (0,0,P_3)$ is a spontaneous polarization vector of uniaxial ferroelectric. $\varepsilon_g$ is the dielectric layer permittivity, $\varepsilon_S$ is the semiconductor background permittivity. Semiconductor layer thickness is regarded much higher than its screening radius $R_d$. For the case of multi-layer graphene this leads to additional limitations on graphene layers numbers, which correlate with the previous ones. Technologically for the case of multi-layer graphene the dielectric layer means dielectric substrate (the most common one is SiO$_2$, although other dielectrics, including ones with high permittivity, can be used as well – see Ref.[45]).

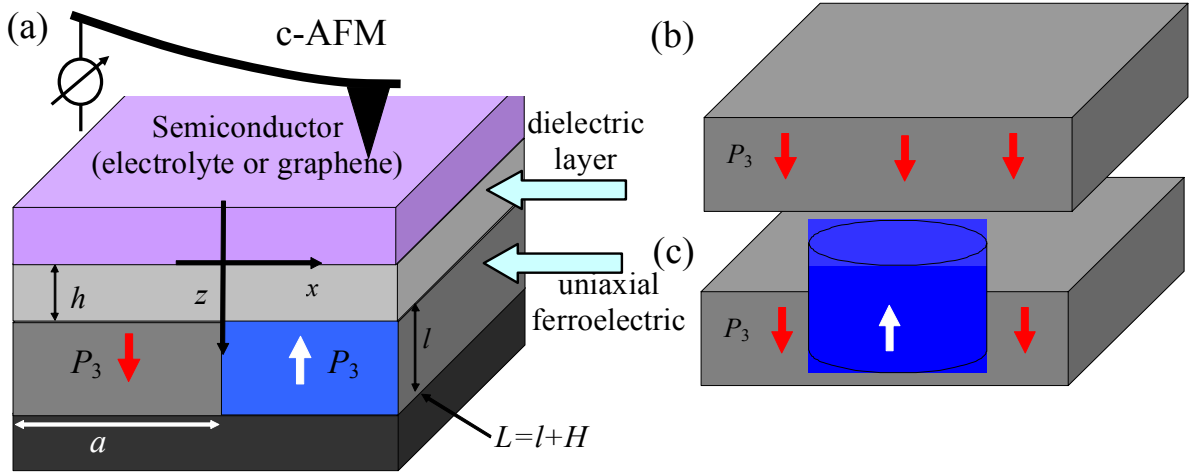

**Figure 1.** Geometry of the considered heterostructure semiconductor/dielectric layer/uniaxial ferroelectric with (a) domain stripes, (b) single-domain, (c) cylindrical domain.

Equation of state $\mathbf{D} = \varepsilon_0 \varepsilon_g \mathbf{E}$ relates the electrical displacement $\mathbf{D}$ and electric field $\mathbf{E}$ in the dielectric layer, $\varepsilon_g$ is the dielectric layer (gap) permittivity, $\varepsilon_0$ is the universal dielectric constant. In ferroelectric film the electric displacement is $\mathbf{D} = \varepsilon_0 \mathbf{E} + \mathbf{P} \approx \varepsilon_0 \hat{\varepsilon}_{ij}^f \mathbf{E} + \mathbf{P}_S$, where $\hat{\varepsilon}_{ij}^f$ is the linear dielectric permittivity tensor. The potential φ of quasi-stationary electric field, $\mathbf{E} = -\nabla \varphi$, can be introduced. The potential $\varphi$ satisfies Laplace's equation inside the dielectric layer. For semiconductor/multilayered graphene we regard the Debye-Hückel approximation validity. This leads to the system of electrostatic equations:

$$\Delta \varphi_S - \frac{\varphi_S}{R_d^2} = 0, \quad \text{for} \quad -\infty < z < 0, \quad \text{(semiconductor)} \quad (1a)$$

$$\Delta \varphi_g = 0, \quad \text{for} \quad 0 < z < h, \quad \text{(dielectric layer)} \quad (1b)$$

$$\left( \varepsilon_{33}^f \frac{\partial^2}{\partial z^2} + \varepsilon_{11}^f \Delta_\perp \right) \varphi_f = 0, \quad \text{for} \quad h < z < L. \quad \text{(ferroelectric film)} \quad (1c)$$

Laplace operator is $\Delta$, $R_d$ is a Debye screening radius. In Equations (1) we used that $\text{div}\mathbf{P}_S(x, y) = 0$ for uncharged 180-degree domain structure. Equations should be supplemented with the boundary conditions of zero potentials at $z << -R_d$ and $z = L$, $\varphi_S(x, y, -\infty) \to 0$ and $\varphi_f(x, y, L) = 0$. Continuous potential on the boundaries between semiconductor and dielectric layer, $\varphi_S(x, y, 0) = \varphi_g(x, y, 0)$, the layer and ferroelectric, $\varphi_g(x, y, h) = \varphi_f(x, y, h)$. Continuous normal component of displacement on the boundaries between semiconductor, namely $-\varepsilon_S(\partial \varphi_S/\partial z) + \varepsilon_g(\partial \varphi_g/\partial z) = 0$ at $z = 0$ and $-\varepsilon_0 \varepsilon_{33}^f (\partial \varphi_f/\partial z) + P_3 + \varepsilon_0 \varepsilon_g (\partial \varphi_g/\partial z) = 0$ at $z = h$.

General solution of Eq.(1) valid for arbitrary distribution $P_3(x, y)$ is derived in the **Appendix A1.** For a single-domain ferroelectric analytical solution of Eqs.(1) has a relatively simple form:

$$\varphi_S(z) = \frac{-P_3 R_d \varepsilon_g l \exp(z/R_d)}{\varepsilon_0 \left( \varepsilon_g \varepsilon_S l + \varepsilon_{33}^f (\varepsilon_S h + \varepsilon_g R_d) \right)} \quad \text{for} \quad -\infty < z < 0, \quad (2a)$$

$$\varphi_g(z) = \frac{-P_3 l (R_d \varepsilon_g + \varepsilon_S z)}{\varepsilon_0 \left( \varepsilon_g \varepsilon_S l + \varepsilon_{33}^f (\varepsilon_S h + \varepsilon_g R_d) \right)} \quad \text{for} \quad 0 < z < h, \quad (2b)$$

$$\varphi_f(z) = \frac{P_3 (\varepsilon_S h + \varepsilon_g R_d)(z - L)}{\varepsilon_0 \left( \varepsilon_g \varepsilon_S l + \varepsilon_{33}^f (\varepsilon_S h + \varepsilon_g R_d) \right)} \quad \text{for} \quad h < z < L. \quad (2c)$$

Here $l = L - h$ is the thickness of ferroelectric film. Expressions (2) are derived in **Appendix A2.** Both in Debye-Hückel approximation the space charge density $\rho_S(z)$ in semiconductor (or multilayered graphene) and the total charge $\sigma_S \approx -\frac{\varepsilon_0 \varepsilon_S}{R_d^2} \int_0^{-\infty} \varphi_S(z) dz$ are given by the elementary expressions:

$$\rho_S(z) = \frac{\varepsilon_0 \varepsilon_S}{R_d^2} \varphi_S(z), \quad (3a)$$

$$\sigma_S = \frac{-P_3 \varepsilon_g \varepsilon_S l}{\varepsilon_g \varepsilon_S l + \varepsilon_{33}^f (\varepsilon_S h + \varepsilon_g R_d)}. \tag{3b}$$

Note, that the total charge is equal to $-P_3$ only in the case of perfect screening, i.e. at $h = 0$ (no gap) and $R_d = 0$ (perfect conductor instead of semiconductor).

Direct analyses of e.g. denominators in expressions (2)-(3) tells us that the effective gap $h^*$ can be introduced as

$$h^* = h + \frac{\varepsilon_g}{\varepsilon_S} R_d \tag{4}$$

It is important for further narration, that the effective gap can be introduced in a simple way (4) only for a single-domain case.

For the case of rectangular-shape 1D-periodic domain stripes with a period $a$, $P_3(x) \approx \sum_{m=0}^{\infty} P_m \sin(k_m x)$, where $P_m \approx 4P_S/\pi(2m+1)$ and $k_m = (2m+1)(2\pi/a)$, the solution of Eqs.(1) have the form of series listed in the **Appendix A3.** Solution for the case of the cylindrical domain with radius $a$ is derived in **Appendix A4.** For the case of domain stripes and cylindrical domains the relation between their size $a$, film thickness $L$, "true" gap $h$ and even $R_d$ becomes to play a decisive role in a possibility to introduce the effective gap $h^*$. Going ahead we would like to state that Eq.(4) remain approximately valid under the condition $(\varepsilon_S h + \varepsilon_g R_d) \ll a$ and $a \gg l$. Under the condition Eqs.(2) describe adequately the potential distribution in the central part of domain(s).

### 3. Electric field, potential and space charge redistribution caused by domain structure

Below we study the electric field, potential and space charge redistribution caused by ferroelectric domain structure for the heterostructure "liquid or solid electrolyte/dead layer/ferroelectric LiNbO$_3$" and "multi-layer graphene/dielectric Al$_2$O$_3$ /ferroelectric Rochelle salt". Material parameters of the heterostructures used in the calculations are listed in the **Table 1.**

**Table 1.** Material parameters of the heterostructures used in the calculations

| | Semiconductor/dielectric/ferroelectric | |
|---|---|---|
| **Parameters** | paraelectric or electrolyte /background dead layer/LiNbO$_3$ | multi-layer graphene/dielectric Al$_2$O$_3$/ Rochelle salt |
| semiconductor screening radius $R_d$ | 5 nm | 1 nm * |
| semiconductor permittivity $\varepsilon_S$ | 80 (liquid or solid electrolyte, or SrRuO$_3$) | 15 |

| dielectric permittivity $\varepsilon_g$ | 5 (background constant) | 12.53 (sapphire) |
|---|---|---|
| dielectric layer thickness $h$ | 1 nm | 5 nm** |
| dielectric anisotropy of ferroelectric $\gamma$ | 0.58 | 3.87 |
| ferroelectric permittivity $\varepsilon_{33}^f$ | 29 | 300 |
| ferroelectric polarization $P_S$ | 0.75 C/m$^2$ | 0.002 C/m$^2$ |
| ferroelectric thickness $l$ | 300 nm | 300 nm |

*calculated using the formulae $R_d = \sqrt{\varepsilon_0 \varepsilon_s k_B T / (2e^2 N_0)}$ from the carrier density $N_0 = 10^{25} \text{m}^{-3}$ at room temperature.
** 7 nm wide Al$_2$O$_3$ substrate for graphene layer was fabricated in [46].

The dependence of electric field lateral ($E_x$) and vertical ($E_z$) components on the $x$ and $z$ coordinates were studied for the heterostructure electrolyte layer/dead layer/ferroelectric LiNbO$_3$ [**Figures 2a,b**] and for the heterostructure multi-layer graphene/dielectric Al$_2$O$_3$/ferroelectric Rochelle salt [**Figures 2c,d**]. As one can see from the **Figs. 2a** and **2c**, both electric field components have relatively small value in semiconductor/multi-layer graphene and they decrease quasi-linearly with increasing the distance from the interface with dielectric. In dielectric, both components are an order of magnitude higher than in the semiconductor/multi-layer graphene; they increase when approach the ferroelectric and have maximum on the interface with ferroelectric, then vanishes in accordance with the exponential law, wherein $E_z$ changes its sign at the boundary between dielectric and ferroelectric, in contrast to the x component which retains its sign. **Figures 2b** and **2d** demonstrates a non-trivial dependence of electric field on the lateral coordinate $x$ originated from the domain modulation. It can be seen that $E_x$ is the even x-function and has only one extremum at the boundary between two domains. In the dielectric gap $E_x$-curve becomes more flattened only. $E_z$ component is the odd function and has more complicated structure than $E_x$, in particular it has one extremum in the semiconductor/multi-layer graphene located at the domain boundary, and two additional local extrema appear near the domain boundaries in the ferroelectric. Note, that the absolute values in the ferroelectric are at least two times bigger than in the semiconductor/multi-layer graphene. In general, the lateral $E_z$-modulation is in-phase, while $E_x$-modulation is anti-phase with the ferroelectric polarization of domains.

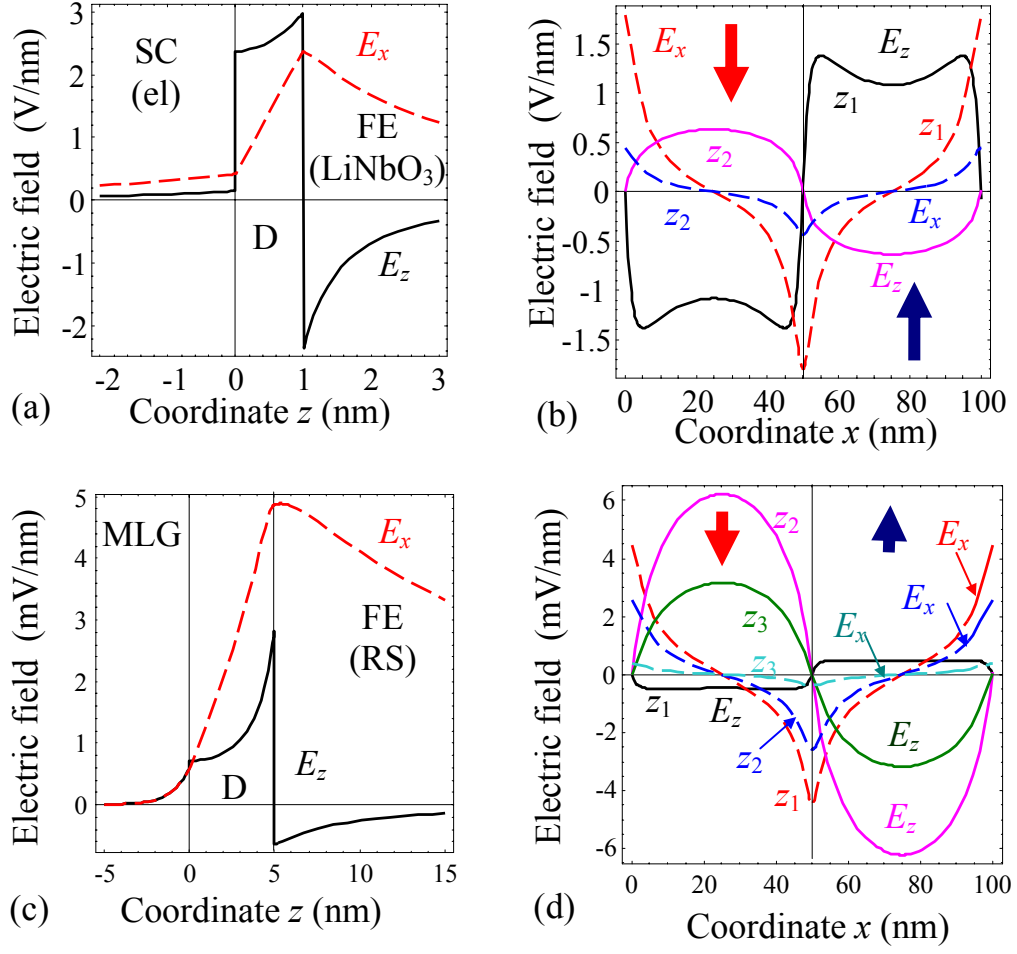

**Figure 2. Electric field distribution caused by domain stripes of period 100 nm.** Profiles are calculated across **(a,c)** and along the hetero-interface **(b,d)**. Symbols SC and MLG indicate the semiconductor or multi-layer graphene respectively, D is dielectric, FE is ferroelectric. For the system **electrolyte/dead layer/LiNbO$_3$** (plots **a, b**) distance x=1nm for plot **(a)** and $z_1$=2nm and $z_2$= −0.5nm for plot **(b)**. For the system **multi-layer graphene/Al$_2$O$_3$/Rochelle salt** (plots **c, d**) distance x=1nm for plot **(c)** and $z_1$=10 nm, $z_2$=2.5 nm and $z_3$= −0.5nm for plot **(d).**

**Figure 3** shows the electric potential distribution along $x$ and $z$ coordinates calculated for the same heterostructures as in the **Figs.2**. Note that the potential curve is symmetric with respect to the sign change of the polarization vector. Potential z-dependence has sharp maximum at the interface between dielectric and ferroelectric. The potential exponentially decreases with the distance z from the interface inside the semiconductor and linearly increases in the dielectric gap (**Fig. 3a,d**). The dependence of potential on $x$-coordinate in ferroelectric is similar to the harmonic function, and becomes more flattened in the semiconductor (**Fig. 3b, e**).

**Figures 3c** and **3f** illustrate the charge density in the semiconductor. The charge densities distributions in semiconductor and multi-layered graphene are similar; both are quasi-sinusoidal functions with zero value at the boundary between two domains and have extremum in the centre of the domain. With moving towards the interface with dielectric, the absolute value of the charge density increases. So that it is evident that ferroelectric polarization of the domains modulates the potential distribution and space-charge density in semiconductor/multi-layer graphene.

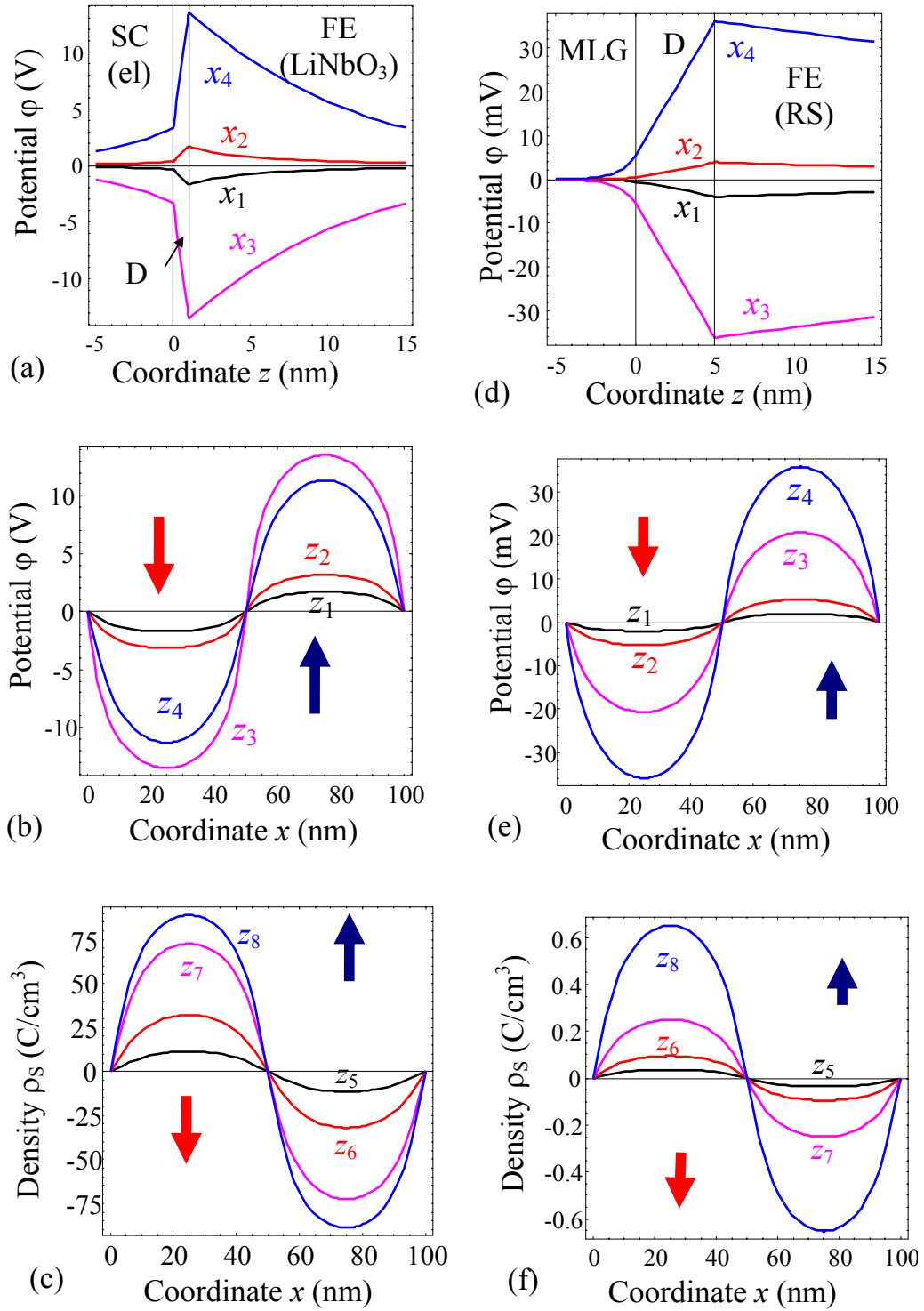

**Figure 3. Electrostatic potential (a,b, d, e), space charge density (c,f) distribution caused by domain stripes of period 100 nm.** Distributions are calculated along and across the interface for

the system **electrolyte/dead layer/LiNbO₃** (plots **a-c**). (**a**) $x_1= -1$, $x_2= 1$, $x_3= -25$, $x_4= +25$ nm; (**b**) $z_1= -3$, $z_2= 0$, $z_3= 1$, $z_4= +3$ nm; (**c**) $z_5= -10$, $z_6= -5$, $z_7= -1$, $z_8= 0$ nm. Plots **d-f** corresponds to the system **multi-layer graphene/Al₂O₃/Rochelle salt**. (**d**) $x_1= -1$, $x_2= 1$, $x_3= -25$, $x_4= +25$ nm; (**e**) $z_1= -1$, $z_2= 0$, $z_3= 2.5$, $z_4= +6$ nm; (**f**) $z_5= -3$, $z_6= -2$, $z_7= -1$, $z_8= 0$ nm.

**Figure 4** shows the electric field $E_x$ and $E_z$ components distributions in dependence on the $x$ and $z$ coordinates, created by a single cylindrical domain. A comparison with electric field caused by domain stripes [shown in the **Figure 2**] underlines several interesting features. For the case of cylindrical domain $E_z$-profile is similar to the one in the case of domain stripes, however its value remains almost constant inside the dielectric layer (at fixed x coordinate) and abruptly changes at the layer boundaries (**Fig.4a,d**). $E_z$ dependence on $x$ coordinate (at fixed $z$ coordinate) has a similar form as it has for the domain stripes case, however in more details it shows the disappearance of local extremums near the domain boundaries with increasing of the distance $z$ from the interface (**Fig.4b,e**). It is worth to underline that the field distribution is very different inside and outside the domain region. Radial component $E_r$ z-dependence has non-trivial form, especially in the ferroelectric part (**Fig.4c,f**). As anticipated from the radial symmetry of the problem the absolute value of $E_r$ is zero in the domain centre ($r=0$), then it increases with the increase of distance $r$ inside the domain; reaches a pronounced maximum at the domain wall and decreases outside the domain.

We can conclude from the **Figure 4** that a cylindrical domain structure of ferroelectric can effectively modulate electric field in the heterostructure, and the modulation of the radial and normal electric field components are in anti-phase to each other. Note that rather strong radial fields can readily cause the spontaneous appearance closure domains if the considered stripe domains were placed in the multiaxial ferroelectric. That is why we consider a uniaxial ferroelectric, where the complications are improbable.

**Figure 5** demonstrates electric potential and free charge redistribution along $x$ and $z$ direction caused by a cylindrical domain. Excluding the fact that these dependencies profile are very different inside and outside the cylindrical domain, they have similar properties as the ones calculated for the domain stripes case. In the rest, all the arguments concerning these dependencies are similar with ones for domain stripes case. In particular, the ferroelectric polarization of a single cylindrical domain creates a drop-like modulation of the electric potential distribution and space-charge density in a semiconductor/multi-layer graphene.

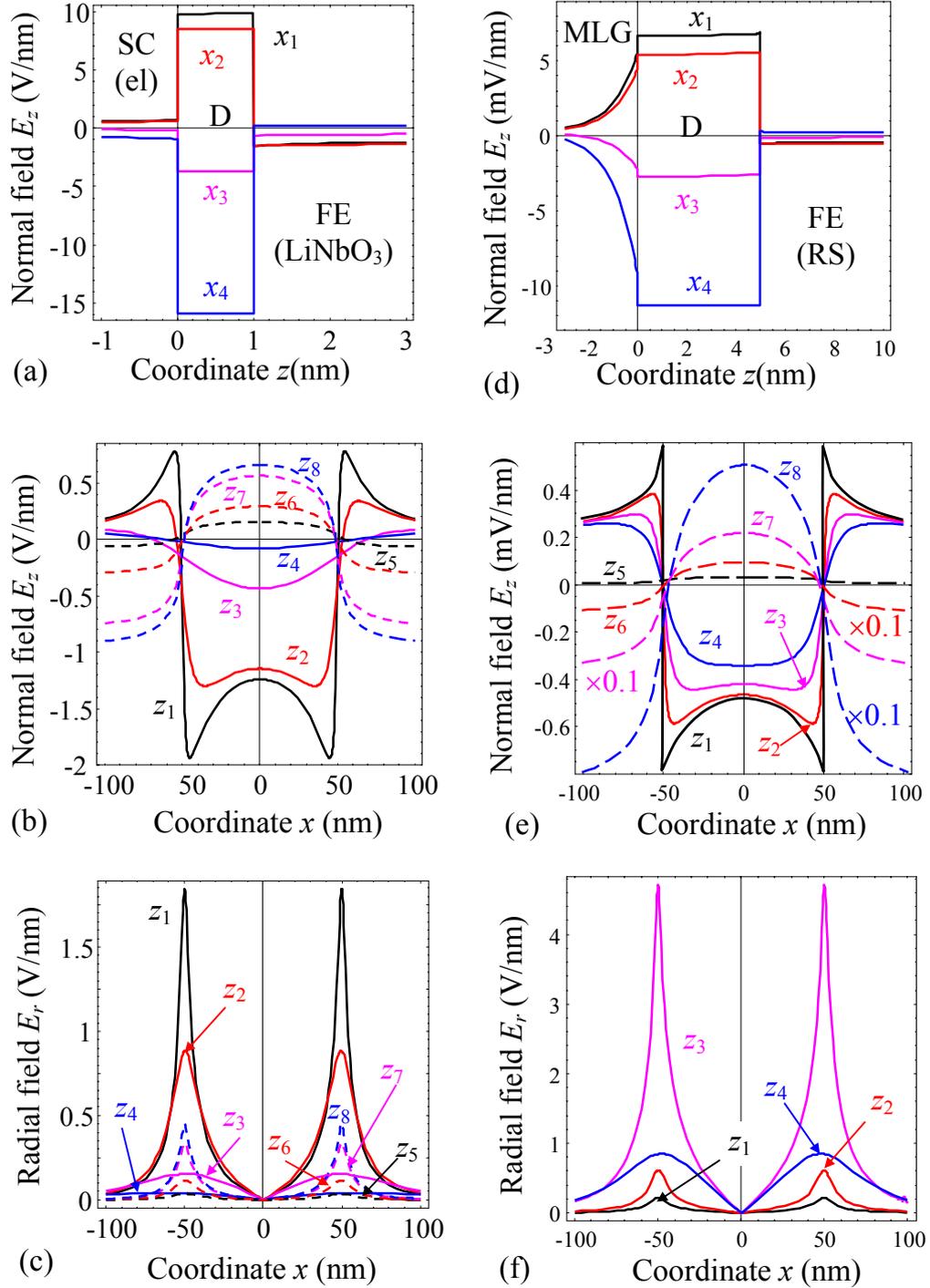

**Figure 4. Electric field components redistribution caused by a cylindrical domain of radius $a$=50 nm.** Plots **a-c** correspond to the **electrolyte/dead layer/ LiNbO$_3$**. (a) $x_1$= 1, $x_2$= 25, $x_3$= 50, $x_4$= 100 nm, **(b, c)** $z_1$= 2, $z_2$= 5, $z_3$= 25, $z_4$= 50 nm (in ferroelectric) and $z_5$= −10, $z_6$= −5, $z_7$= −1, $z_8$= −0.1 nm (in the electrolyte). Plots **d-f** correspond to the **multi-layer graphene/Al$_2$O$_3$/Rochelle salt. (d)** $x_1$= 0, $x_2$= 25, $x_3$= 50, $x_4$= 100 nm, **(e)** $z_1$= 5, $z_2$= 10, $z_3$= 25, $z_4$= 50 nm (in ferroelectric) and $z_5$= −4, $z_6$= −2, $z_7$= −1, $z_8$= −0.1 nm (in graphene). **(f)** $z_1$= −1, $z_2$= 0, $z_3$= 10, $z_4$= 100 nm.

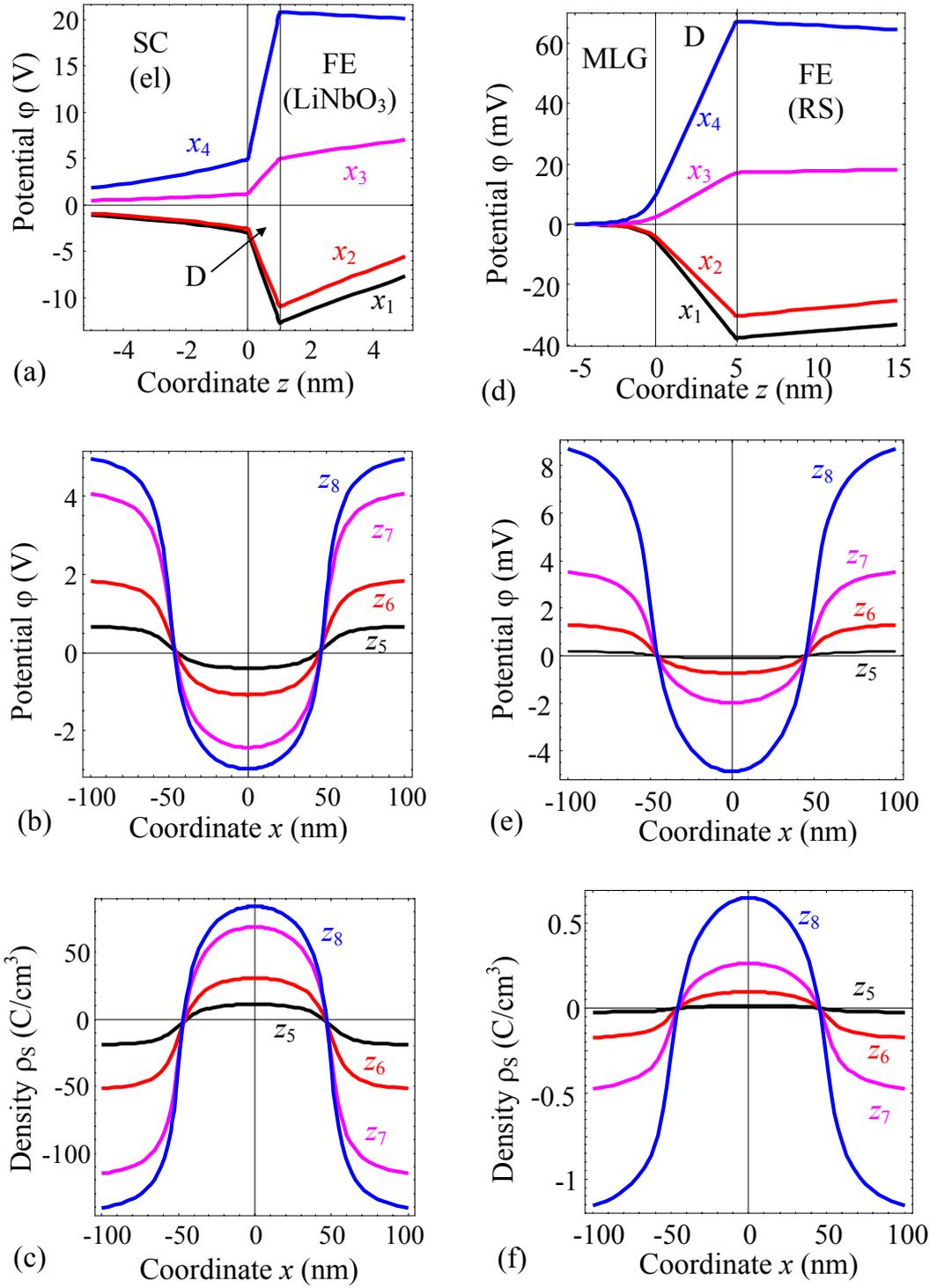

**Figure 5. Electrostatic potential (a,b) and space charge (c,d) redistribution caused by a cylindrical domain of radius $a$=50 nm.** Plots **a-c** correspond to the **electrolyte/dead layer/LiNbO$_3$**. **(a)** $x_1$= 1, $x_2$= 25, $x_3$= 50, $x_4$= 100 nm, **(b, c)** $z_5$= −10, $z_6$= −5, $z_7$= −1, $z_8$= 0 nm. Plots **d-f** correspond to the **multi-layer graphene /Al$_2$O$_3$/Rochelle salt. (d)** $x_1$= 0, $x_2$= 25, $x_3$= 50, $x_4$= 100 nm, **(e,f)** $z_5$= −4, $z_6$= −2, $z_7$= −1, $z_8$= −0.1 nm (in graphene).

# 4. Size effects of the space charge accumulation

Finally let us illustrate the finite size effects of the space charge accumulation in semiconductor caused by a ferroelectric domain structure allowing the electric field drop in the dielectric gap. We demonstrate the size effects on the example of the heterostructures "semiconductor/thin dielectric layer/ferroelectric".

**Figure 6** shows the free space charge dependency on the domain stripe period $a$, dielectric layer thickness $h$ and screening radius $R_d$ calculated for the heterostructure "electrolyte/dead layer/LiNbO$_3$". **Figure7** illustrates the same quantities calculated for the heterostructure" multi-layer graphene/Al$_2$O$_3$/Rochelle salt". In general, charge density is the nonlinear, monotonic function of two variables $a$ and $h$, and the dependence on the variable $a$ is stronger than that on the $h$. One can note, that there is saturation region for the large values of $a$, which is the more pronounced for the lower values of $h$. Charge dependence on the radius $R_d$ is also a nonlinear, monotonic function that has saturation region in range of the small $R_d$ values. Note also, that saturation region becomes wider with increasing of the period $a$. Under the absence of the dielectric layer ($h=0$) the space charge amplitude slows saturates to a spontaneous polarization value with the domain size increase. When the dielectric layer is present, the amplitude saturates to the smaller values, at that the saturation value essentially decreases with the layer thickness increase, but the saturation rate is almost independent on the thickness in the actual range of parameters. The space charge amplitude strongly vanishes with the dielectric layer thickness increase only for small domain period, but remains almost constant for wide domains in accordance with Eq.(3b) [see the first line in the **Table 2**]..

**Figure 8** shows the free space charge dependence on the cylindrical domain radius $a$, dielectric layer thickness $h$ and screening radius $R_d$ calculated for the heterostructure "electrolyte/dead layer/LiNbO$_3$". **Figure 9** shows the same calculated for the heterostructure "multi-layer graphene/Al$_2$O$_3$/Rochelle salt". Charge density is a nonlinear monotonic function of the period $a$ (similar to the stripe domain case), but it is no more a positive function. For small domain period the charge density may have an extremum depending on the dielectric thickness $h$ or depending on the screening radius $R_d$. Note that charge density function has more non-trivial dependency on stripe period, gap thickness and screening radius in the case of cylindrical domain [see the second line in the **Table 2**].

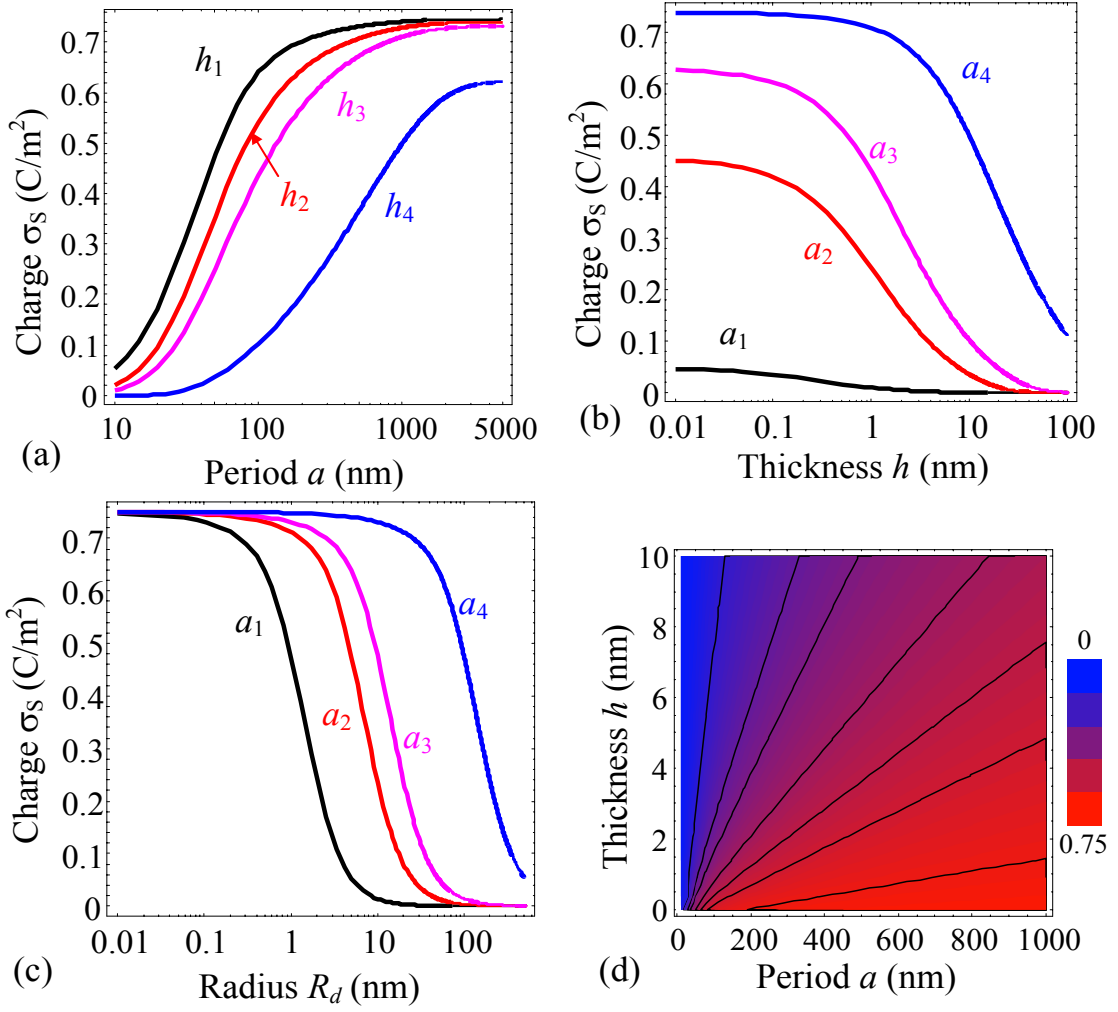

**Figure 6. Size effects of the space charge accumulated by domain stripes.** Plots correspond to the **electrolyte/dead layer/LiNbO$_3$**. **(a)** Maximal total charge of the semiconductor vs. the stripe period $a$ calculated for the screening radius $R_d$=5 nm and different gap thickness $h_1$=0, $h_2$=0.4, $h_3$=1, $h_4$=10 nm. **(b)** Total charge vs. $h$ calculated for $R_d$=5 nm and period $a_1$=10, $a_2$=100, $a_3$=500, $a_4$=1000 nm. **(c)** Total charge vs. $R_d$, calculated for $h$=0, $a_1$=10, $a_2$=50, $a_3$=100, $a_4$=1000 nm. **(d)** Contour map of the total charge in coordinates $\{a,h\}$.

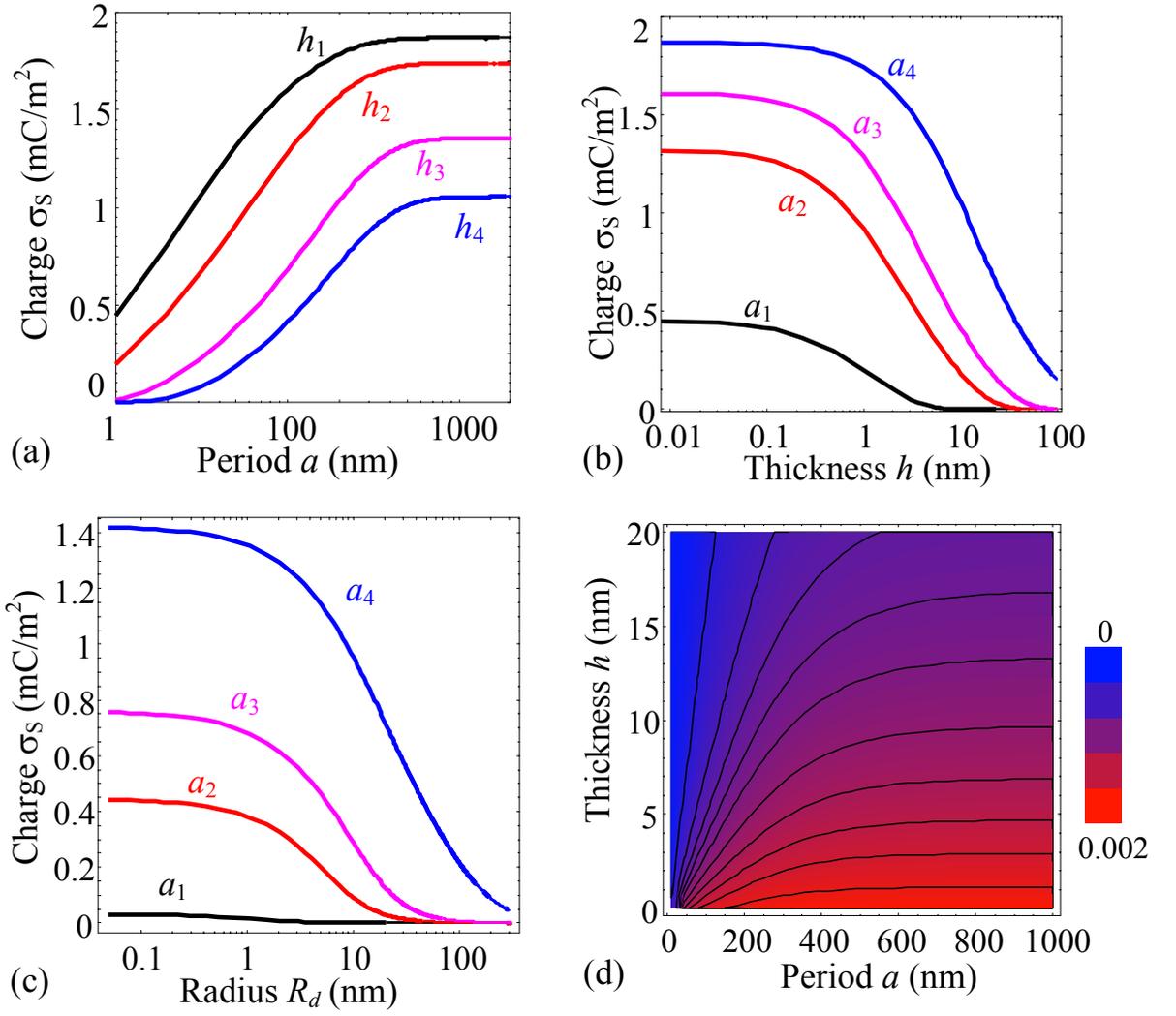

**Figure 7. Size effects of the space charge accumulated by domain stripes.** Plots correspond to the **multi-layer graphene/Al$_2$O$_3$/Rochelle salt. (a)** Maximal total charge of the semiconductor vs. the stripe period $a$ calculated for the screening radius $R_d$=5 nm and different gap thickness $h_1$=0, $h_2$=1, $h_3$=5, $h_4$=10 nm. **(b)** Total charge vs. $h$ calculated for $R_d$=1 nm and period $a_1$=10, $a_2$=50, $a_3$=100, $a_4$=1000 nm. **(c)** Total charge vs. $R_d$, calculated for $h$=5 nm, $a_1$=10, $a_2$=50, $a_3$=100, $a_4$=1000 nm. **(d)** Contour map of the total charge in coordinates $\{a,h\}$.

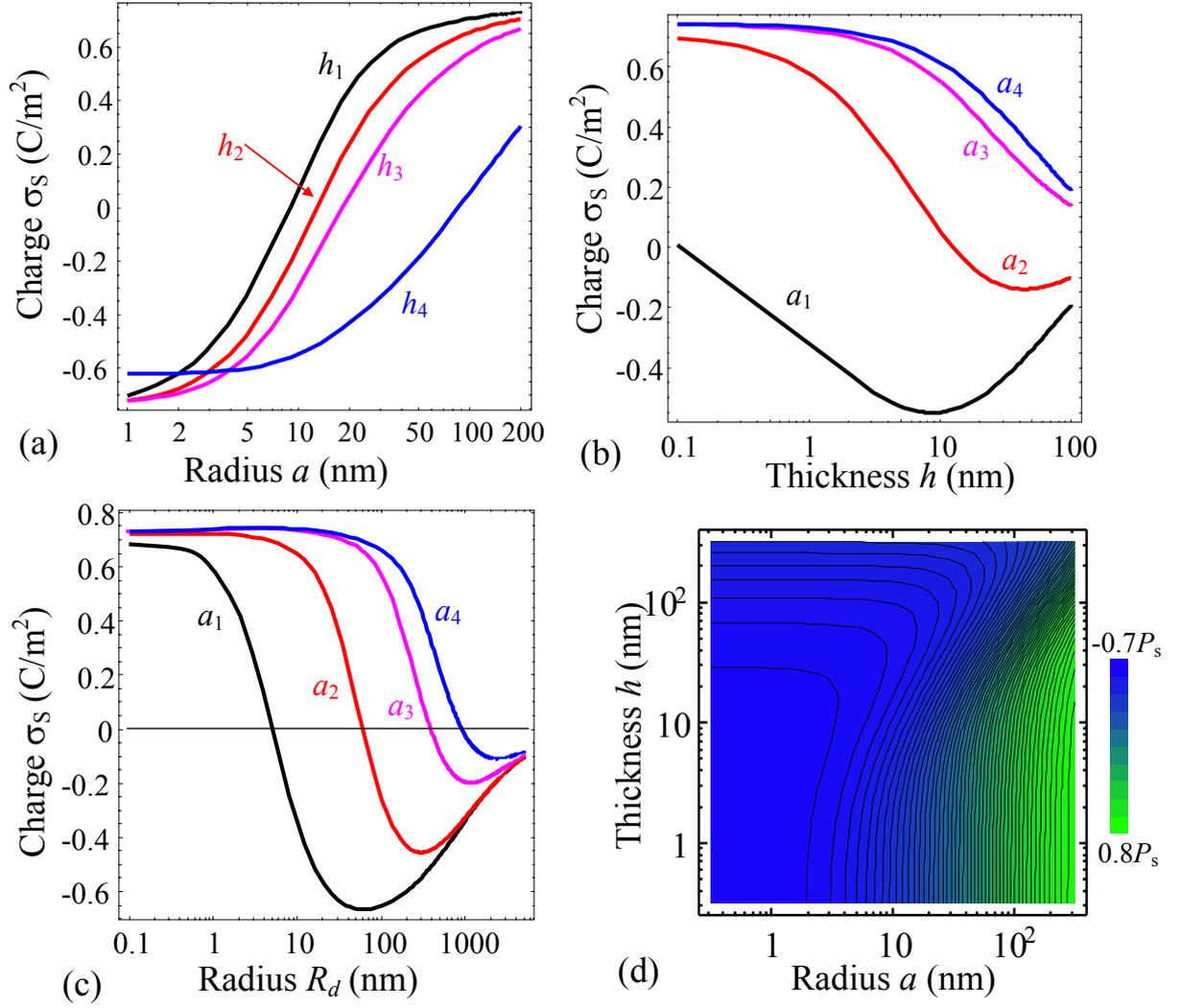

**Figure 8. Size effects of the space charge accumulated by a cylindrical domain.** Plots correspond to the **electrolyte/dead layer/LiNbO$_3$**. **(a)** The total charge of the semiconductor vs. the domain radius $a$ calculated in the point $x = y = 0$ for the screening radius $R_d$=5 nm and different gap thickness $h_1$=0, $h_2$=0.4, $h_3$=1, $h_4$=10 nm. **(b)** Total charge vs. $h$ calculated for $R_d$=5 nm and radii $a_1$=10, $a_2$=100, $a_3$=500, $a_4$=1000 nm. **(c)** Total charge vs. $R_d$, calculated for $h$=0, $a_1$=10, $a_2$=50, $a_3$=100, $a_4$=1000 nm. **(d)** Contour map of the total charge in coordinates $\{a,h\}$.

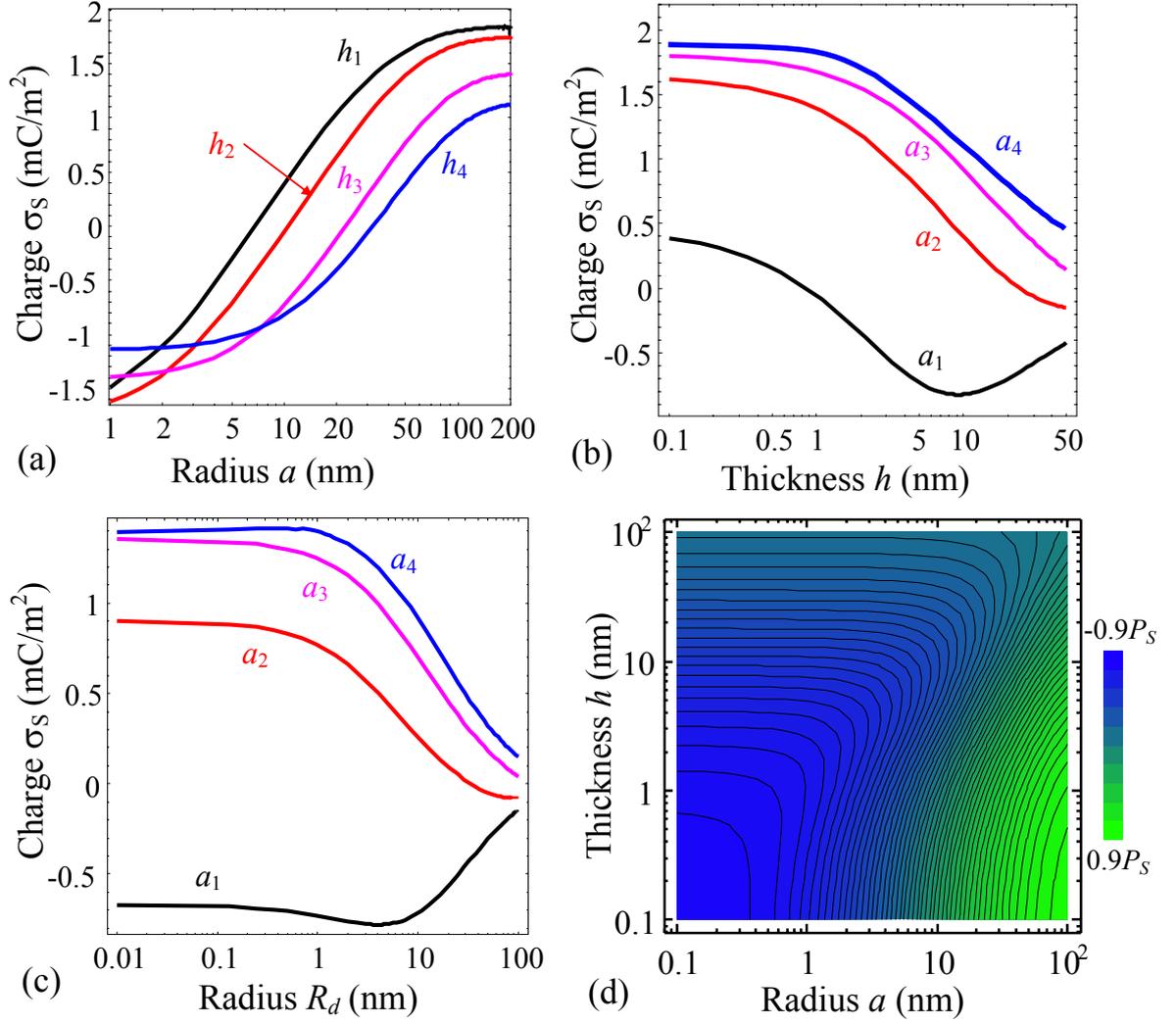

**Figure 9. Size effects of the space charge accumulated by a cylindrical domain.** Plots correspond to the **multi-layer graphene/ Al$_2$O$_3$/Rochelle salt. (a)** The total charge of the semiconductor vs. the domain radius $a$ calculated in the point $x = y = 0$ for the screening radius $R_d$=1 nm and different gap thickness $h_1$=0, $h_2$=1, $h_3$=5, $h_4$=10 nm. **(b)** Total charge vs. $h$ calculated for $R_d$=1 nm and radii $a_1$=10, $a_2$=100, $a_3$=500, $a_4$=1000 nm. **(c)** Total charge vs. $R_d$, calculated for $h$= 5nm, $a_1$=10, $a_2$=50, $a_3$=100, $a_4$=1000 nm. **(d)** Contour map of the total charge in coordinates $\{a,h\}$.

**Table 2. Size effect of the charge density modulation $\sigma_S$**

|  | Parameter | | |
| --- | --- | --- | --- |
|  | **Dielectric layer thickness $h$** | **Screening radius $R_d$** | **Domain size $a$** |
| **Charge density modulation $\sigma_S$ induced by stripe domains** | Monotonically decreases with $h$ increase tends to zero in the limit | Monotonically decreases with $R_d$ increase and tends to zero in the limit $R_d/a \to \infty$ | Monotonically increases with $a$ increase and slowly saturates to the |

| | $h/a \to \infty$ | | maximal value |
|---|---|---|---|
| **Charge density $\sigma_S$ variation induced by a cylindrical domain** | Monotonically decreases with $h$ increase for $a >> h$. Has a minima at $a \sim h$. | Decreases with $R_d$ increase, reached a minimum, that is especially pronounced for $a \sim h$, and tends to zero in the limit $R_d/a \to \infty$ | $\dfrac{-P_3}{1+\left(\varepsilon_{33}^f h^*/\varepsilon_g \varepsilon_S l\right)}$ in the limit $a/h^* \to \infty$, here $h^* = \varepsilon_S h + \varepsilon_g R_d$ (in agreement with Eq.(3b)) |

## 5. Concluding remarks

We consider a heterostructure "domain patterned ferroelectric film – ultra-thin dielectric layer (or physical gap) – semiconductor", where the semiconductor can be an incipient ferroelectric, electrolyte or multi-layered graphene. Unexpectedly we have found that the space charge modulation profile and amplitude in the semiconductor, that screens the spontaneous polarization of a 180-degree domain structure of ferroelectric, nonlinearly depends on the domain structure period $a$, dielectric layer thickness $h$ and semiconductor screening radius $R_d$ in a rather non-trivial way. Multiple size effects appearance and manifestation are defined by the relationship between the $a$, $h$ and $R_d$. Under the absence of the dielectric layer the space charge amplitude very slows saturates to a spontaneous polarization value with the domain size increase. When the dielectric layer is present, the amplitude saturates to the smaller values, at that the saturation value essentially decreases with the layer thickness increase, but the saturation rate is almost independent on the thickness in the actual range of parameters. The space charge amplitude strongly vanishes with the dielectric thickness increase only for small domain period, but remains almost constant for wide domains. We show that the concept of effective gap $h^*$ can be introduced in a simple way only for a single-domain limit. Depth distribution of electric potential is nonlinear inside ferroelectric, while it is linear inside the dielectric and exponentially vanishes inside the semiconductor as anticipated. Obtained analytical results open the way for understanding of current-AFM results for contaminated ferroelectric surfaces in ambient atmosphere with moderate humidity, since c-AFM contrast can be regarded proportional to the relative electron density.

## Acknowledgements

A.N.M. and E.A.E. acknowledge the support via bilateral SFFR-NSF project (US National Science Foundation under NSF-DMR-1210588 and State Fund of Fundamental Research of Ukraine, grant UU48/002). M.V.S. acknowledges State Fund of Fundamental Research of Ukraine, grant 53.2/006. S.V.K. acknowledges Office of Basic Energy Sciences, U.S. Department of Energy. Y.H.C acknowledge the National Science Council, R.O.C. (NSC-101-2119-M-009-003-MY2), Ministry of Education (MOE-ATU 101W961), and Center for Interdisciplinary Science of National Chiao Tung University. V.Y.S. acknowledges the Russian Foundation of Basic Research (grant 14-02-92709 Ind-a).

2119-M-009-003-MY2), Ministry of Education (MOE-ATU 101W961), and Center for Interdisciplinary Science of National Chiao Tung University. V.Y.S. acknowledges the Russian Foundation of Basic Research (grant 14-02-92709 Ind-a).

# Supplemental Materials to
# Ferroelectric domain triggers the charge modulation in semiconductors

### Appendix A. Electric field calculations in thermodynamic equilibrium
**A1. General case of 2D-domain structure**

Equations of state relate electrical displacement **D** and electric field **E** in the dielectric gap as

$$\mathbf{D}_g = \varepsilon_0 \varepsilon_g \mathbf{E}_g, \qquad (A.1)$$

In ferroelectric the electric displacement is:

$$\mathbf{D}_f = \varepsilon_0 \mathbf{E}_f + \mathbf{P} \approx \varepsilon_0 \hat{\varepsilon}_{ij}^f \mathbf{E}_f + \mathbf{P}_S(x,y,z). \qquad (A.2)$$

Here $\mathbf{P}(x,y,z)$ is polarization vector, $\mathbf{P}_S(x,z) = (0,0,P_3(x,z))$ is spontaneous polarization vector. $\varepsilon_g$ is the dielectric layer (gap) permittivity, $\varepsilon_0$ is the universal dielectric constant.

Electrostatic quasi-stationary Maxwell equation rot **E**=0 should be valid in the actual frequency range, giving one the opportunity to introduce the potential φ of quasi-stationary electric field, $\mathbf{E}_{g,f}(x,y,z,t) = -\nabla \varphi_{g,f}(x,y,z,t)$. Inside the dielectric gap potential φ satisfies Laplace's equation. For semiconductor we regard the Debye approximation validity. Allowing for these assumptions and Eqs.(A.1)-(A.2), electrostatic potential satisfy the system of equations:

$$\left(\frac{\partial^2}{\partial z^2} + \frac{\partial^2}{\partial x^2} + \frac{\partial^2}{\partial y^2}\right)\varphi_S - \frac{\varphi_S}{R_d^2} = 0, \quad \text{for } -\infty < z < 0, \text{ (semiconductor)} \quad (A.3a)$$

$$\left(\frac{\partial^2}{\partial z^2} + \frac{\partial^2}{\partial x^2} + \frac{\partial^2}{\partial y^2}\right)\varphi_g = 0, \quad \text{for } 0 < z < h, \text{ (dielectric)} \quad (A.3b)$$

$$\left(\varepsilon_{33}^f \frac{\partial^2}{\partial z^2} + \varepsilon_{11}^f \left(\frac{\partial^2}{\partial x^2} + \frac{\partial^2}{\partial y^2}\right)\right)\varphi_f = 0, \quad \text{for } h < z < L. \text{ (ferroelectric)} \quad (A.3c)$$

In Eq.(A.3) we used that $\text{div}\mathbf{P}_S(x,y) = 0$ for uncharged 180-degree domain structure. Eqs.(A.3) should be supplemented with the boundary conditions of zero potentials at $z \ll -R_d$ and $z = L$, continuous potential and normal component of displacement on the boundaries between semiconductor, dielectric gap and ferroelectric, namely

$$\varphi_S(x,y,z \ll -R_d) \to 0 \qquad (A.4a)$$

$$\varphi_S(x,y,0) = \varphi_g(x,y,0), \qquad (A.4b)$$

$$D_{Sn}(x,y,0) - D_{gn}(x,y,0) \equiv \varepsilon_0 \left(-\varepsilon_S \frac{\partial \varphi_S(x,y,0)}{\partial z} + \varepsilon_g \frac{\partial \varphi_g(x,y,0)}{\partial z}\right) = 0, \qquad (A.4c)$$

$$\varphi_g(x,y,h) = \varphi_f(x,y,h), \qquad (A.4d)$$

$$D_{fn} - D_{gn} \equiv -\varepsilon_{33}^{f} \frac{\partial \varphi_{f}(x,y,h)}{\partial z} + \frac{P_{3}(x,y)}{\varepsilon_{0}} + \varepsilon_{g} \frac{\partial \varphi_{g}(x,y,h)}{\partial z} = 0, \quad \text{(A.4e)}$$

$$\varphi_{f}(x, y, z = L) = 0. \quad \text{(A.4f)}$$

Let us look for the general solution of Eqs.(A.3) in the form for $-\infty < z < 0$:

$$\varphi_{S}(x,y,z) = \int_{-\infty}^{\infty} dk_{x} \int_{-\infty}^{\infty} dk_{y} \varphi_{0}(k_{x},k_{y}) \exp\left(ik_{x}x + ik_{y}y + z\sqrt{k^{2} + \frac{1}{R_{d}^{2}}}\right), \quad \text{(A.5a)}$$

for $h < z < L$:

$$\varphi_{g}(x,y,z) = \int_{-\infty}^{\infty} dk_{x} \int_{-\infty}^{\infty} dk_{y} (\varphi_{1}(k_{x},k_{y})\exp(-zk) + \varphi_{2}(k_{x},k_{y})\exp(zk)) \exp(ik_{x}x + ik_{y}y), \quad \text{(A.5b)}$$

$$\varphi_{f}(x,y,z) = \int_{-\infty}^{\infty} dk_{x} \int_{-\infty}^{\infty} dk_{y} \varphi_{3}(k_{x},k_{y}) \sinh\left((L-z)\frac{k}{\gamma}\right) \exp(ik_{x}x + ik_{y}y) \quad \text{(A.5c)}$$

Here $\gamma = \sqrt{\varepsilon_{33}^{f}/\varepsilon_{11}^{f}}$ is the dielectric anisotropy factor and $k = \sqrt{k_{x}^{2} + k_{y}^{2}}$. Solution (A.5a) identically satisfies the boundary conditions (A.4a). Solution (A.5c) identically satisfies the boundary condition (A.4f). Substitution of the solution (A.5) into the boundary conditions (A.4b,c,d,e) leads to the system of four equations for the four functions $\varphi_{i}(k)$ ($i$=0, 1, 2, 3):

$$\varphi_{0} = \varphi_{1} + \varphi_{2}, \quad \text{(A.6a)}$$

$$-\varepsilon_{S} \sqrt{k^{2} + \frac{1}{R_{d}^{2}}} \varphi_{0} + \varepsilon_{g} k(-\varphi_{1} + \varphi_{2}) = 0, \quad \text{(A.6b)}$$

$$\varphi_{1} \exp(-kh) + \varphi_{2} \exp(kh) = \varphi_{3} \sinh\left((L-h)\frac{k}{\gamma}\right), \quad \text{(A.6c)}$$

$$\varepsilon_{33}^{f} \frac{k}{\gamma} \varphi_{3} \cosh\left((L-h)\frac{k}{\gamma}\right) + \frac{\tilde{P}_{3}}{\varepsilon_{0}} + \varepsilon_{g} k(-\varphi_{1}\exp(-kh) + \varphi_{2}\exp(kh)) = 0, \quad \text{(A.6d)}$$

where $P_{3}(x,y) = \int_{-\infty}^{\infty} dk_{x} \int_{-\infty}^{\infty} dk_{y} \tilde{P}_{3}(k_{x},k_{y}) \exp(ik_{x}x + ik_{y}y)$ is regarded known. The solution of the system (A.6) with respect to $\varphi_{i}(k)$ has the form:

$$\varphi_{0} = -\frac{\varepsilon_{g} \tilde{P}_{3}(k)\gamma}{\varepsilon_{0} Det(k)} \sinh\left(l\frac{k}{\gamma}\right) \quad \text{(A.7a)}$$

$$\varphi_{1} = \frac{-\tilde{P}_{3}(k)\gamma}{2\varepsilon_{0} k Det(k)} \left(\varepsilon_{g} k - \varepsilon_{S} \sqrt{k^{2} + \frac{1}{R_{d}^{2}}}\right) \sinh\left(l\frac{k}{\gamma}\right) \quad \text{(A.7b)}$$

$$\varphi_{2} = \frac{-\tilde{P}_{3}(k)\gamma}{2\varepsilon_{0} k Det(k)} \left(\varepsilon_{g} k + \varepsilon_{S} \sqrt{k^{2} + \frac{1}{R_{d}^{2}}}\right) \sinh\left(l\frac{k}{\gamma}\right) \quad \text{(A.7c)}$$

$$\varphi_3 = \frac{-\tilde{P}_3(k)\gamma}{\varepsilon_0 k Det(k)} \left( \varepsilon_g k \cosh(kh) + \varepsilon_S \sqrt{k^2 + \frac{1}{R_d^2}} \sinh(kh) \right) \quad \text{(A.7d)}$$

$$Det(k) = \begin{pmatrix} \varepsilon_{33}^f \cosh\left(l\frac{k}{\gamma}\right)\left(\varepsilon_g k \cosh(kh) + \varepsilon_S \sqrt{k^2 + \frac{1}{R_d^2}} \sinh(kh)\right) + \\ + \varepsilon_g \gamma \left( \varepsilon_S \sqrt{k^2 + \frac{1}{R_d^2}} \cosh(kh) + \varepsilon_g k \sinh(kh) \right) \sinh\left(l\frac{k}{\gamma}\right) \end{pmatrix} \quad \text{(A.7e)}$$

Hereinafter $l = L - h$ is the true thickness of ferroelectric film.

In accordance Boltzmann statistics, hole density is $p(x,y,z) \approx N_0 \exp\left(-\frac{e\varphi_S(x,y,z)}{k_B T}\right)$ and electron density is $n(x,y,z) \approx N_0 \exp\left(\frac{e\varphi_S(x,y,z)}{k_B T}\right)$ in a proper semiconductor, so the charge density and the total sheet charge in semiconductor are:

$$\rho_S(x,y,z) = e(p(x,y,z) - n(x,y,z)) = -2N_0 e \sinh\left(\frac{e\varphi_S(x,y,z)}{k_B T}\right) \approx -\frac{\varepsilon_0 \varepsilon_S}{R_d^2} \varphi_S(x,y,z), \quad \text{(A.8a)}$$

$$\sigma_S(x,y) = \int_{-\infty}^{0} \rho_S(x,y,z)dz \approx -\frac{\varepsilon_0 \varepsilon_S}{R_d^2} \int_{-\infty}^{0} \varphi_S(x,y,z)dz. \quad \text{(A.8b)}$$

In Debye approximation, $\left|\frac{e\varphi}{k_B T}\right| \ll 1$, the density $\rho_S(x,y,z) \approx -\frac{\varepsilon_0 \varepsilon_S}{R_d^2} \varphi_S(x,y,z)$, as anticipated, since $R_d = \sqrt{\frac{\varepsilon_0 \varepsilon_S k_B T}{2e^2 N^0}}$. The distribution the screening charge $\rho_S(x)$ modulates the conductance $\sigma(x)$ of the semiconductor layer. Within the simplest Drude model: $\sigma(x) = e\mu\sigma_S(x)$, here $e$ is the electron charge, $\mu$ is the mobility.

## A2. One-dimensional periodic domain structure

Using the fact, that a periodic 1D-domain structure of conventional rectangular shape and period $a$ can be expanded in Fourier series as

$$P_3(x) \approx \sum_{m=0}^{\infty} P_m \sin(k_m x), \quad P_m \approx \frac{4P_S}{(2m+1)\pi}, \quad k_m = \frac{2\pi}{a}(2m+1). \quad \text{(A.9a)}$$

Corresponding 1D-Fourier image is

$$\tilde{P}_3(k_x) \propto \sum_m \frac{P_m}{2i} (\delta(k_x - k_m) - \delta(k_x + k_m)) \quad \text{(A.9b)}$$

Solution (A.5) acquires the form

$$\varphi_S(x,z) = \sum_m \varphi_0(k_m)\exp\left(z\sqrt{k_m^2 + \frac{1}{R_d^2}}\right)\sin(k_m x), \quad \text{for} \quad -\infty < z < 0, \quad \text{(A.10a)}$$

$$\varphi_g(x,z) = \sum_m (\varphi_1(k_m)\exp(-z|k_m|) + \varphi_2(k_m)\exp(z|k_m|))\sin(k_m x), \quad \text{for} \quad 0 < z < h, \quad \text{(A.10b)}$$

$$\varphi_f(x,z) = \sum_m \varphi_3(k_m)\sinh\left((L-z)\frac{|k_m|}{\gamma}\right)\sin(k_m x) \quad \text{for} \quad h < z < L. \quad \text{(A.10c)}$$

In Equations (A.10) one should substitute $\tilde{P}_3(k) \to P_m$ in Eqs.(A.7), e.g.

$$\varphi_0(k_m) = -\frac{\varepsilon_g P_m \gamma}{\varepsilon_0 Det(k_m)}\sinh\left(l\frac{k_m}{\gamma}\right).$$

In Debye approximation the density $\rho_S(x,z) \approx -\frac{\varepsilon_0 \varepsilon_S}{R_d^2}\varphi_S(x,z)$ and the total charge is

$$\sigma_S(x,y) = -\frac{\varepsilon_0 \varepsilon_S}{R_d^2}\sum_m \frac{\varphi_0(k_m)\sin(k_m x)}{\sqrt{k_m^2 + R_d^{-2}}}. \quad \text{(A.10d)}$$

Equation (A.10d) in explicit form gives $\sigma_S(x) = \frac{\varepsilon_0 \varepsilon_S}{R_d^2}\sum_m \frac{\varepsilon_g P_m \gamma}{\varepsilon_0 Det(k_m)}\sinh\left(l\frac{k_m}{\gamma}\right)\frac{\sin(k_m x)}{\sqrt{k_m^2 + R_d^{-2}}}$, where

$P_m \approx \frac{4P_S}{(2m+1)\pi}$, $k_m = \frac{2\pi}{a}(2m+1)$. In the limit $2\pi h/a \ll 1$, $\varepsilon_{33}^f \gg \gamma\varepsilon_g$ and $l/a \gg 1$ we get

$$\sigma_S(x) \approx \frac{(4/\pi)\gamma\varepsilon_S\varepsilon_g P_S \sin(kx)}{R_d^2\sqrt{k^2 + R_d^{-2}}\left(\varepsilon_{33}^f k\left(\varepsilon_g + \varepsilon_S h\sqrt{k^2 + R_d^{-2}}\right) + \varepsilon_g \gamma\left(\varepsilon_S\sqrt{k^2 + R_d^{-2}} + \varepsilon_g h k^2\right)\right)} \propto \frac{(4/\pi)\gamma\varepsilon_S\varepsilon_g P_S \sin(kx)}{\varepsilon_{33}^f R_d^2 k\left(\varepsilon_g\sqrt{k^2 + R_d^{-2}} + \varepsilon_S h(k^2 + R_d^{-2})\right)},$$

where $k = 2\pi/a$.

**A3. Single-domain case**

Expression for the electric potential in the case, when the ferroelectric is single-domain, can be obtained by tending the wave vector $k$ to the zero value. As result of such approach, electric potential is given by the following expressions.

$$\varphi_S(z) = \frac{-P_3 R_d \varepsilon_g l \exp(z/R_d)}{\varepsilon_0\left(\varepsilon_g\varepsilon_S l + \varepsilon_{33}^f(\varepsilon_S h + \varepsilon_g R_d)\right)} \quad \text{for} \quad -\infty < z < 0, \quad \text{(A.11a)}$$

$$\varphi_g(z) = \frac{-P_3 l(R_d\varepsilon_g + \varepsilon_S z)}{\varepsilon_0\left(\varepsilon_g\varepsilon_S l + \varepsilon_{33}^f(\varepsilon_S h + \varepsilon_g R_d)\right)} \quad \text{for} \quad 0 < z < h, \quad \text{(A.11b)}$$

$$\varphi_f(z) = \frac{P_3(\varepsilon_S h + \varepsilon_g R_d)(z - L)}{\varepsilon_0\left(\varepsilon_g\varepsilon_S l + \varepsilon_{33}^f(\varepsilon_S h + \varepsilon_g R_d)\right)} \quad \text{for} \quad h < z < L. \quad \text{(A.11c)}$$

Corresponding charge density (A.8) is given by the elementary expressions:

$$\rho_S(z) = -2N_0 e \sinh\left(\frac{e\varphi_S(z)}{k_B T}\right) \approx -\frac{\varepsilon_0 \varepsilon_S}{R_d^2} \varphi_S(z), \quad (A.12a)$$

$$\sigma_S = \int_{-\infty}^{0} \rho_S(x,y,z)dz \approx +\frac{\varepsilon_0 \varepsilon_S}{R_d^2} \int_{0}^{-\infty} \varphi_S(z)dz = \frac{-P_3 \varepsilon_g \varepsilon_S l}{\varepsilon_g \varepsilon_S l + \varepsilon_{33}^f (\varepsilon_S h + \varepsilon_g R_d)}. \quad (A.12b)$$

Finally, let us study the limit $R_d \to 0$ in Eqs.(A.11). In the limit $\varphi_S(z) = 0$ and

$$\varphi_g(z) = \frac{-P_3 l z}{\varepsilon_0 (\varepsilon_g l + \varepsilon_{33}^f h)}, \quad \varphi_f(z) = \frac{P_3(z-L)h}{\varepsilon_0 (\varepsilon_g l + \varepsilon_{33}^f h)}, \quad \sigma_S = \frac{-P_3 \varepsilon_g l}{\varepsilon_g l + \varepsilon_{33}^f h}. \quad (A.13)$$

Direct comparison of the last relatively simple expression (A.13) with Eqs.(A.12b) tells us that the effective gap

$$h^* = h + \frac{\varepsilon_g}{\varepsilon_S} R_d \quad (A.14)$$

can be introduced in order to describe the charge density in semiconductor. $l = L - h$ is the true thickness of ferroelectric film.

Electric field distribution

$$E_S(z) = \frac{P_3 \varepsilon_g l}{\varepsilon_0 (\varepsilon_g \varepsilon_S l + \varepsilon_{33}^f (\varepsilon_S h + \varepsilon_g R_d))} \exp(z/R_d) \quad \text{for} \quad -\infty < z < 0, \quad (A.15a)$$

$$E_g(z) = \frac{P_3 l \varepsilon_S}{\varepsilon_0 (\varepsilon_g \varepsilon_S l + \varepsilon_{33}^f (\varepsilon_S h + \varepsilon_g R_d))} \quad \text{for} \quad 0 < z < h, \quad (A.15b)$$

$$E_f(z) = \frac{-P_3 (\varepsilon_S h + \varepsilon_g R_d)}{\varepsilon_0 (\varepsilon_g \varepsilon_S l + \varepsilon_{33}^f (\varepsilon_S h + \varepsilon_g R_d))} \quad \text{for} \quad h < z < L. \quad (A.15c)$$

Electrostatic displacement in ferroelectric media is

$$D_f(z) = \frac{P_3 \varepsilon_g \varepsilon_S l}{(\varepsilon_g \varepsilon_S l + \varepsilon_{33}^f (\varepsilon_S h + \varepsilon_g R_d))} \quad (A.16)$$

**A4. Cylindrical domain in ferroelectric film**

Let us consider a single cylindrical domain with sharp domain wall with polarization profile,

$$P_3(x,y) = P_S (2\theta(a-r)-1), \quad (A.15a)$$

where, $r = \sqrt{x^2 + y^2}$, $\theta(x)$ is a Hevisaid step-function, $a$ is the domain radius. Fourier image corresponding to a single cylindrical domain is:

$$\tilde{P}_S(k_x, k_y) = 2P_S \left( a \frac{J_1(ka)}{k} - \pi\delta(k_x)\delta(k_y) \right) \quad (A.15b)$$

Where $k = \sqrt{k_x^2 + k_y^2}$, $J_1(x)$ is a Bessel function of the first order. Solution (A.5) acquires the form:

$$\varphi_S(r,z) = \int_0^\infty dk\, k J_0(kr) \varphi_0(k) \exp\left(z\sqrt{k^2 + \frac{1}{R_d^2}}\right), \qquad (A.16a)$$

for $h < z < L$:

$$\varphi_g(r,z) = \int_0^\infty dk\, k J_0(kr)(\varphi_1(k)\exp(-zk) + \varphi_2(k)\exp(zk)), \qquad (A.16b)$$

$$\varphi_f(r,z) = \int_0^\infty dk\, k J_0(kr) \varphi_3(k) \sinh\left((L-z)\frac{k}{\gamma}\right) \qquad (A.16c)$$

Here $J_0(x)$ is a Bessel function of zero order, kernels $\varphi_i(k)$ are given by Eqs.(A.7). The charge density is $\rho_S(r,z) \approx -\frac{\varepsilon_0 \varepsilon_S}{R_d^2} \varphi_S(r,z)$ in accordance with Eq.(A.8) and the total sheet charge in semiconductor:

$$\sigma_S(r) = \int_{-\infty}^0 \rho_S(r,z)dz \approx -\frac{\varepsilon_0 \varepsilon_S}{R_d^2} \int_{-\infty}^0 \varphi_S(r,z)dz . \qquad (A.17)$$